# Capacity Limits for the Cognitive Radio Channel with Confidential Messages

Reza K. Farsani, Reza Ebrahimpour

Emails: reza_khosravi@alum.sharif.ir, ebrahimpour@ipm.ir

School of Cognitive Sciences, Institute for Research in Fundamental Sciences (IPM), P.O. Box 19395-5746, Tehran, Iran

***Abstract*—As a brain inspired wireless communication scheme, *cognitive radio* is a novel approach to promote the efficient use of the scarce radio spectrum by allowing some users called "cognitive users" to access the under-utilized spectrum licensed out to the "primary users". Besides highly reliable communication and efficient utilization of the radio spectrum, the security of information transmission against eavesdropping is critical in the cognitive radios for many potential applications. In this paper, this problem is investigated from an information theoretic viewpoint. Capacity limits are explored for the Cognitive Radio Channel (CRC) with confidential messages. As an idealized information theoretic model for the cognitive radio, this channel includes two transmitters which send independent messages to their corresponding receivers such that one transmitter, i.e., the *cognitive transmitter*, has access non-causally to the message of the other transmitter, i.e., the *primary transmitter*. The message designated to each receiver is required to be kept confidential with respect to the other receiver. The secrecy level for each message is evaluated using the *equivocation rate*. Novel inner and outer bounds for the capacity-equivocation region are established. It is shown that these bounds coincide for some special cases. Specifically, the capacity-equivocation region is derived for a class of less-noisy CRCs and also a class of semi-deterministic CRCs. For the case where only the message of the cognitive transmitter is required to be kept confidential, the capacity-equivocation region is also established for the Gaussian CRC with weak interference.**

***Keywords*-Cognitive Radio Channel; Information Theoritic Security; Capacity-Equivocation Region.**

## I. INTRODUCTION

One of the main challenges in today's wireless communication systems is how to efficiently manage the use of the electromagnetic radio spectrum as a precious yet limited natural resource. The radio spectrum is licensed by governments to be used by transmitters and receivers in communication networks and demands for its utilization are continually increasing. According to official reports [1], the frequency bands allocated to many users remain unused most of the time. Therefore, the utilization of the radio spectrum can be considerably improved by allowing a secondary user (cognitive user) to access the frequency bound unoccupied by the primary user. The cognitive radio is a novel approach proposed for this purpose [2]. It is an intelligent wireless communication system which is capable of sensing the environment and adapting its parameters to statistical variations in the input stimuli with the purpose of filling voids in the wireless spectrum [2]. The main objectives of this brain-inspired wireless communication scheme are to achieve highly reliable communication and efficient utilization of the spectrum [2-4]. Therefore, diverse areas of research- from learning theory and cognitive sciences to communication theory and signal processing- are engaged to successfully develop such intelligent communication technology. The study of the cognitive networks is of great importance in communications due to their role in the fifth generation of wireless communication systems [3], motivating a growing body of research which has been dedicated to this issue in recent years. However, there still exist many open problems regarding the limits of these networks. Specifically, similar to any other communication network, a fundamental problem is to determine capacity limits for the cognitive radio, i.e., all communication rates at which information can be reliably transmitted over the system. This problem is basically analyzed in information theory. The fundamental limits of communications derived in information theory provide criteria for the operation of the cognitive networks so that "*researchers may gauge the efficiency of any practical network as well as draw inspiration as to which direction to pursue in their design*" [5]. For a more detailed discussion on the motivation behind information theoretic study of cognitive radio networks, please refer to [6] and [7]. From the information theoretic viewpoint, the cognitive radio was initially considered in [7, 8]. The authors of [7] proposed a general idealized model for a *Cognitive Radio Channel* (CRC). In this channel model, two transmitters send independent messages to their corresponding receivers such that one of the transmitters named the *cognitive transmitter* has access non-causally[1] to the message of the other transmitter named the *primary transmitter*. An achievable rate region was also derived in [7] for this channel. Following up [7], capacity limits have been widely studied for the CRC. In particular, various capacity inner and outer bounds and also some partial capacity results have been derived for the channel in [8-12]. For a comprehensive review of the existing results, please refer to the recent work [13] and the literature therein. However, the capacity of the CRC in the general case is still an open problem.

Similar to other multi-user communication networks, the security of information transmission against eavesdropping is critical in the cognitive radios for many potential applications. The scope of this paper is to investigate this problem from an information theoretic viewpoint. Information theoretic security was indeed pioneered by Shannon [14]. Later in a significant paper [15], Wyner introduced the notion of *wiretap channel* that is a communication scenario where a transmitter sends a confidential message to a legitimate receiver while being eavesdropped by a wire-tapper (which is assumed to have no computational limitation). Wyner characterized the capacity-equivocation rate region for the degraded wire-tapper channel where the wire-tapper receives a degraded version of the legitimate receiver signal. As given in [15, 16], the equivocation rate indicates a portion of the message rate which can be transferred to the legitimate receiver, while the wire-tapper is kept completely incognizant of this part. The outstanding consequence of Wyner's work is that (for the degraded channel) the transmitter can send information to the legitimate receiver in perfect secret at a positive rate. In other words, from the information theoretic point of view, secure communication can be achieved by coding at the physical layer (while in cryptography, secrecy is achieved by data encryption at the application layer). The result of Wyner was later generalized by Csiszar and Korner [16] to the two-user broadcast channel (not necessarily degraded) with a private message for one of the users and a common message for both users. In recent years, information theoretic security has been widely studied for different multi-user networks (for example, see [17] and literature therein). Specifically, bounds for the secrecy capacity region of the two-user Broadcast Channel (BC) and the two-user Classical Interference Channel (CIC) are derived in [18]. The two-user BC with a common and two confidential private messages is studied in [19] where inner and outer bounds for the capacity-equivocation region are given. The capacity-equivocation region is also established for the less-noisy and the semi-deterministic BCs. In [20], a full characterization is established for the capacity-equivocation region of a cognitive interference channel. The difference between the model of [20] and the CRC introduced in [7] is that, in the former the message of the primary transmitter is decoded at both receivers and only the message of the cognitive transmitter is required to be kept confidential from the non-corresponding receiver. The cognitive interference channel model of [20] is recently re-considered in [21] where the authors investigate security under randomness constraint for the system. The cognitive interference channel with two confidential messages was also considered in [22]. In [23], an achievable rate region is given for the CRC where only the cognitive message is required to be kept confidential. However, the bounds derived in the latter two papers are not efficient to prove any capacity result for the channel.

[1] In fact, in this idealized model it is assumed that a genie provides non-causal knowledge of the primary transmitter's message to the cognitive transmitter. To see that why this is a viable model to explore, please refer to [6], (see also [7]).

In this paper, we explore capacity limits for the CRC with confidential messages where the message designated to each receiver is required to be kept confidential with respect to the other receiver. The secrecy level for each message is evaluated by the equivocation rate. We establish novel inner and outer bounds for the capacity-equivocation region of the CRC. Our outer bound has an efficient structure; it includes certain constraints that are not given in any of the outer bounds derived in previous papers [9-12, 22] for the CRC. As will be shown, these constraints enable us to prove important capacity results for the channel. Our inner bound is also obtained by a novel approach. In [18] and [19], to derive achievable rate-equivocation region for the two-user broadcast channel the authors make use of a rather sophisticated technique called *double binning* (a bin of sub-bins of random codewords is generated to encode each confidential message and hence the name of bauble binning). This technique combines two phases of binning: one for joint precoding as in the Marton's scheme for the broadcast channel without secrecy [24] and the other for preserving of confidentiality of messages. The double binning technique requires a complicated scheme for codeword assigning to messages (see for example [19, page 4537]). In this paper, to derive our inner bound for the CRC with confidential messages, instead of double binning, we present a new scheme that requires only a single phase of binning. In our scheme the sizes of the bins are intelligently designed to simultaneously guarantee both joint precoding and preserving of confidentiality. The benefit of our approach is that it requires a simple scheme for codeword assigning to messages.

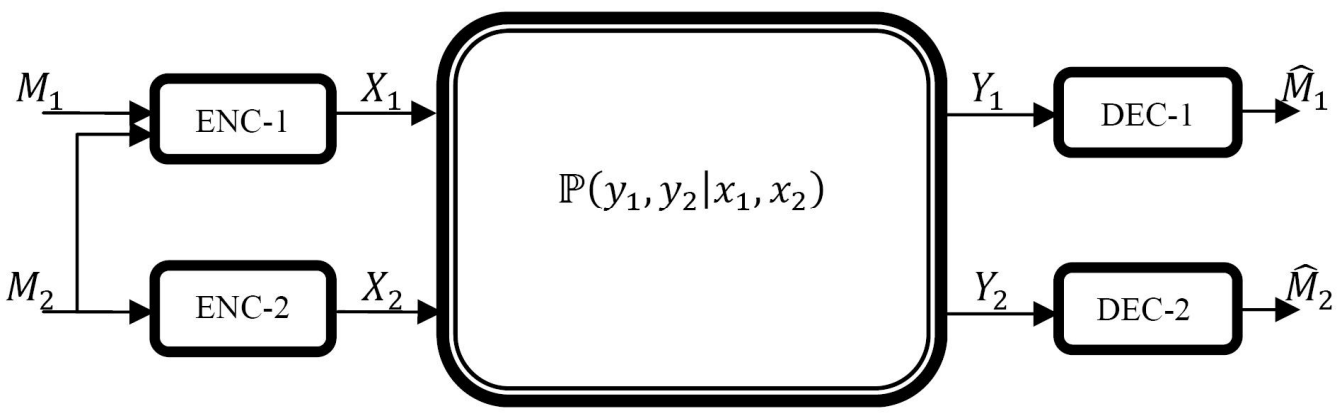

Figure 1. The CRC: $X_1$ is the cognitive transmitter, $X_2$ is the primary transmitter, $Y_1$ is the cognitive receiver and $Y_2$ is the primary receiver.

We then show that our derived inner and outer bounds yield exact capacity for some special cases. Specifically, the capacity-equivocation region is established for a class of less-noisy CRCs and also a class of semi-deterministic CRCs. For the case where only the message of the cognitive transmitter is required to be kept confidential, we also obtain the capacity-equivocation region for the Gaussian CRC with weak interference. Thus, this paper establishes the first capacity results for the CRC with confidential messages. These results are important as the main problem from the information theoretic point of view is to establish capacity limits of a given network. Once these limits are known, one can proceed to design practical coding schemes that achieve them.

We provide preliminaries and channel model definitions in Section II. The main results are given in Section III. We finally present our concluding remarks. Appendix includes some of the long proofs.

## II. Preliminaries and Definitions

In this paper, we use the following notations: Random Variables (RV) are denoted by upper case letters (e.g. $X$) and lower case letters are used to show their realization (e.g. $x$). The range set of $X$ is represented by $\mathcal{X}$. The set of all jointly $\epsilon$-letter (strongly) typical $n$-sequences $(x^n, y^n)$ with respect to the PDF $P_{XY}(x,y)$ is denoted by $\mathcal{T}_\epsilon^n(P_{XY})$, (see [25, Sec. 1.5]). Also, for a given sequence $x^n$, the notation $\mathcal{T}_\epsilon^n(P_{XY}|x^n)$ is used to denote the set of all $n$-sequences $y^n$ such that $(x^n, y^n) \in \mathcal{T}_\epsilon^n(P_{XY})$. The set of nonnegative real numbers is denoted by $\mathbb{R}_+$. Given a real number $x$, the function $[x]_+$ is equal to $x$ if $x$ is positive and zero otherwise. Finally, the function $\psi(x)$ is defined as: $\psi(x) \equiv \frac{1}{2}\log(1+x)$, for $x \in \mathbb{R}_+$.

***Definition 1:*** A discrete memoryless CRC denoted by $\{\mathcal{X}_1, \mathcal{X}_2, \mathcal{Y}_1, \mathcal{Y}_2, \mathbb{P}(y_1, y_2|x_1, x_2)\}$ is a channel which is organized by two input alphabet sets $\mathcal{X}_1, \mathcal{X}_2$, two output alphabet sets $\mathcal{Y}_1, \mathcal{Y}_2$, and a transition probability function $\mathbb{P}(y_1, y_2|x_1, x_2)$ that describes the relation between the inputs and outputs of the channel. Fig. 1 illustrates a model of the channel.

The channel is assumed to be memoryless, i.e.,

$$P(y_1^n, y_2^n|x_1^n, x_2^n) = \prod_{t=1}^{n} \mathbb{P}(y_{1,t}, y_{2,t}|x_{1,t}, x_{2,t}), \qquad n \geq 1$$

The Gaussian channel is given by the following standard form:

$$\begin{cases} Y_1 = X_1 + aX_2 + Z_1 \\ Y_2 = bX_1 + X_2 + Z_2 \end{cases} \tag{1}$$

where $Z_1, Z_2$ are zero-mean unit-variance Gaussian RVs and $\mathbb{E}[X_i^2] \leq P_i, i = 1,2$.

For the CRC shown in Fig. 1, given a natural number $n$ and two real numbers $R_1$ and $R_2$, a length-$n$ code $\mathfrak{C}^n(R_1, R_2)$ with two private messages $M_1$ and $M_2$ uniformly distributed over the sets $\{1, \dots, 2^{nR_1}\}$ and $\{1, \dots, 2^{nR_2}\}$, respectively, consists of the following:

1. Two encoding functions that are given by:

$$\begin{cases}\xi_1(.):\{1,\dots,2^{nR_1}\}\times\{1,\dots,2^{nR_2}\}\to\mathcal{X}_1^n\\ \xi_2(.):\{1,\dots,2^{nR_2}\}\to\mathcal{X}_2^n\end{cases}$$

where the generated codewords at the cognitive and the primary transmitters are as $X_1^n=\xi_1(M_1,M_2)$ and $X_2^n=\xi_2(M_2)$, respectively. We consider the case where the cognitive transmitter applies a stochastic encoder ($\xi_1(.)$ defines a transition probability distribution $\xi_1(x_1^n|m_1,m_2)$ where $x_1^n\in\mathcal{X}_1^n$ , $m_1\in\{1,\dots,2^{nR_1}\}$ , and $m_2\in\{1,\dots,2^{nR_2}\}$), while the primary transmitter applies a deterministic encoder ($\xi_2(.)$ is a deterministic function).

2. Two decoder functions that are given by:

$$\mathfrak{D}_i:\mathcal{Y}_i^n\to\{1,\dots,2^{nR_i}\},\qquad i=1,2$$

where the message $M_i$ is estimated as $\widehat{M}_i=\mathfrak{D}_i(Y_i^n)$ at the receiver $Y_i$.

For a given code $\mathfrak{C}^n(R_1,R_2)$, the secrecy level of the confidential messages at the non-corresponding receivers is measured by the normalized equivocation given below:

$$\begin{cases}R_{e,1}^{(n)}\triangleq\frac{1}{n}H(M_1|Y_2^n)\\ R_{e,2}^{(n)}\triangleq\frac{1}{n}H(M_2|Y_1^n)\end{cases}$$

Also, the average error probability of the code, denoted by $P_e^{\mathfrak{C}^n}$, is defined as:

$$P_e^{\mathfrak{C}^n}\triangleq P\left(\bigcup_{i=1,2}\{\mathfrak{D}_i(Y_i^n)\neq M_i\}\right)$$

***Definition 2:*** For the CRC with confidential messages shown in Fig. 1, a rate-equivocation quadruple $(R_1,R_2,R_{e,1},R_{e,2})\in\mathbb{R}_+^4$ is said to be achievable if there exists a sequence of codes $\mathfrak{C}^n(R_1,R_2)$ with:

$$\begin{cases}\lim_{n\to\infty}P_e^{\mathfrak{C}^n}=0\\ R_{e,i}\leq\liminf_{n\to\infty}R_{e,i}^{(n)},\qquad i=1,2\end{cases}$$

***Definition 3:*** The *capacity-equivocation region* of the CRC is the closure of the set of all achievable quadruples $(R_1,R_2,R_{e,1},R_{e,2})$.

For a given message, it is said that perfect secrecy is achieved when its communication rate is equal to the respective equivocation rate, i.e., $R_i=R_{e,i}$, for $i=1$ or $2$. Accordingly, the secrecy capacity region is defined as follows.

***Definition 4:*** The *secrecy capacity region* of the CRC is the closure of the set of all rate pairs $(R_1,R_2)$ so that the 4-tuple $(R_1,R_2,R_1,R_2)$ belongs to the capacity-equivocation region.

Our main results are given in the next section.

## III. Main Results

We begin by establishing a novel capacity outer bound for the CRC with confidential messages. It is given in the following theorem.

***Theorem 1)*** *Define the rate region $\mathfrak{R}_o^{UVW}$ as follows:*

$$\mathfrak{R}_o^{UVW}\triangleq\bigcup_{\substack{P_{WUVX_1X_2},\\ W\to U\to X_1,X_2\\ W\to V\to X_1,X_2}}\left\{\begin{array}{l}(R_1,R_2,R_{e,1},R_{e,2})\in\mathbb{R}_+^4:\\ R_{e,1}\leq R_1,\qquad R_{e,2}\leq R_2,\\ R_1\leq\min\left\{\begin{array}{c}I(U;Y_1),I(U,V;Y_1|X_2),\\ I(U;Y_1|W)+I(W;Y_2),\\ I(U;Y_1|V,X_2)+I(V;Y_2|X_2)\end{array}\right\}\\ R_2\leq\min\left\{\begin{array}{c}I(V,X_2;Y_2),\\ I(V,X_2;Y_2|W)+I(W;Y_1)\end{array}\right\},\\ R_1+R_2\leq I(U;Y_1|V,X_2)+I(V,X_2;Y_2)\\ R_1+R_2\leq I(V,X_2;Y_2|U)+I(U;Y_1)\\ R_1+R_2\leq I(U;Y_1|V,X_2)+I(V,X_2;Y_2|W)\\ \qquad\qquad+I(W;Y_1)\\ R_1+R_2\leq I(V,X_2;Y_2|U)+I(U;Y_1|W)\\ \qquad\qquad+I(W;Y_2)\\ ,\\ R_{e,1}\leq[I(U;Y_1|W)-I(U;Y_2|W)]_+\\ R_{e,1}\leq[I(U;Y_1|V,X_2)-I(U;Y_2|V,X_2)]_+\\ R_{e,2}\leq[I(V,X_2;Y_2|W)-I(V,X_2;Y_1|W)]_+\\ R_{e,2}\leq[I(V,X_2;Y_2|U)-I(V,X_2;Y_1|U)]_+\end{array}\right\}\qquad(2)$$

*The set $\mathfrak{R}_o^{UVW}$ constitutes an outer bound on the capacity-equivocation region for the CRC in Fig. 1.*

The proof of Theorem 1 is given in Appendix. The outer bound $\mathfrak{R}_o^{UVW}$ in (2) indeed has an efficient structure; all capacity results given in this paper are established based on this outer bound. Let us first present some insightful corollaries that are directly deduced from our outer bound.

***Corollary 1)*** *Consider the CRC in Fig. 1 with confidential messages. If the channel satisfies the following condition:*

$$I(U;Y_1|X_2)\leq I(U;Y_2|X_2)\qquad(3)$$

*for all joint PDFs $P_{UX_1X_2}(u,x_1,x_2)$, no secrecy can be achieved for the message $M_1$. Similarly, if the channel satisfies the condition below:*

$$I(V,X_2;Y_2)\leq I(V,X_2;Y_1)\qquad(4)$$

*for all joint PDFs $P_{VX_1X_2}(v,x_1,x_2)$, no secrecy can be achieved for the message $M_2$.*

*Proof of Corollary 1)* First, consider the inequality (3). We prove that it can be extended as follows:

$$I(U;Y_1|V,X_2)\leq I(U;Y_2|V,X_2)\qquad(5)$$

for all joint PDFs $P_{UVX_1X_2}(u,v,x_1,x_2)$. To this end, given a joint PDF $P_{UVX_1X_2}(u,v,x_1,x_2)$, one can write:

$$\begin{aligned}I(U;Y_1|V,X_2)&=\textstyle\sum_v P_V(v)I(U;Y_1|V=v,X_2)\\ &=\textstyle\sum_v P_V(v)I(U;Y_1|X_2)_{\langle P_{UX_1X_2|v}\rangle}\\ &\overset{(a)}{\leq}\textstyle\sum_v P_V(v)I(U;Y_2|X_2)_{\langle P_{UX_1X_2|v}\rangle}\\ &=\textstyle\sum_v P_V(v)I(U;Y_2|V=v,X_2)\\ &=I(U;Y_2|V,X_2)\end{aligned}$$

where the notation $I(A;B|C)_{\langle P(.)\rangle}$ indicates that the mutual information function $I(A;B|C)$ is evaluated by the distribution $P(.)$. Note that for any given $v \in \mathcal{V}$, the function $P_{UX_1X_2|v}$ is a probability distribution defined over the set $\mathcal{U} \times \mathcal{X}_1 \times \mathcal{X}_2$. The inequality (a) is due to (3). Similarly, one can show that the inequality (4) extends as:

$$I(V,X_2;Y_2|W) \le I(V,X_2;Y_1|W) \tag{6}$$

for all joint PDFs $P_{WVX_1X_2}(w,v,x_1,x_2)$. Now, considering (5) and (6), the desired results are directly derived by the constraints on $R_{e,1}$ and $R_{e,2}$ in the outer bound (2). ∎

***Corollary 2)*** For the Gaussian CRC (1), the condition (3) is equivalent to $|b| \ge 1$. Therefore, if $|b| \ge 1$, no secrecy can be achieved for the message $M_1$. Also, if $ab = 1$ and $|b| \le 1 < |a|$, then the Gaussian channel is stochastically degraded ($Y_2$ is a noisy version of $Y_1$) and (4) is satisfied; in this case, no secrecy can be achieved for the message $M_2$.

Now using the outer bound $\mathfrak{R}_o^{UVW}$ in (2), we characterize the capacity-equivocation region for some important special cases of the CRC with confidential messages. First, let us consider the channels which satisfy the condition (4). We refer to these channels as the *less-noisy* CRCs. In the next theorem, we establish the capacity-equivocation region for such channels.

***Theorem 2)*** *Consider the CRC with confidential messages as shown in Fig. 1. The capacity-equivocation region of the less-noisy channel (4) is given by:*

$$\bigcup_{P_{UVX_1X_2}} \left\{ \begin{array}{l} (R_1,R_2,R_{e,1},R_{e,2}) \in \mathbb{R}_+^4 : \\ R_{e,1} \le R_1, \qquad R_{e,2} = 0 \\ R_1 \le I(U,V;Y_1|X_2) \\ R_2 \le I(V,X_2;Y_2) \\ R_1 + R_2 \le I(U;Y_1|V,X_2) + I(V,X_2;Y_2) \\ R_{e,1} \le [I(U;Y_1|V,X_2) - I(U;Y_2|V,X_2)]_+ \end{array} \right\} \tag{7}$$

*Proof of Theorem 2)* The achievability scheme is exactly identical to the one given in [20] for the cognitive interference channel, i.e., a scenario similar to the CRC in Fig. 1 but with the difference that both messages are required at the cognitive receiver. This achievability scheme, as described in details in [20], is based on the superposition coding technique and results in the following rate region:

$$\bigcup_{P_{\tilde{U}VX_2}P_{X_1|\tilde{U}}} \left\{ \begin{array}{l} (R_1,R_2,R_{e,1},R_{e,2}) \in \mathbb{R}_+^4 : \\ R_{e,1} \le R_1, \qquad R_{e,2} = 0 \\ R_1 \le I(\tilde{U},V;Y_1|X_2) \\ R_2 \le \min\{I(V,X_2;Y_1), I(V,X_2;Y_2)\} \\ R_1 + R_2 \le \min\{I(V,X_2;Y_1), I(V,X_2;Y_2)\} \\ \qquad\qquad + I(\tilde{U};Y_1|V,X_2) \\ R_{e,1} \le [I(\tilde{U};Y_1|V,X_2) - I(\tilde{U};Y_2|V,X_2)]_+ \end{array} \right\} \tag{8}$$

Note that since in this coding scheme the message $M_2$ is required at both receivers, its equivocation rate is equal to zero. The details of the proof are omitted as it is given in [20]. By applying the condition (4) to (8) as well as by re-defining $\tilde{U}$ as $\tilde{U} \cong (U,V,X_2)$, we derive the rate region (7). The converse part is directly derived from the outer bound $\mathfrak{R}_o^{UVW}$ given in (2). Note that, due to the less-noisy condition (4), no secrecy is achieved for the message $M_2$. This completes the proof. ∎

***Remark 1:*** Theorem 2 generalizes the result of [19, Th. 3] for the less-noisy BC to the less-noisy CRC in (4). In fact, by setting $X_2 \equiv \emptyset$, Theorem 2 is reduced to [19, Th. 3] specialized for the two-user BC without common message. Also, by setting $R_{e,1} = 0$ in (7), one can easily check that the resultant region is optimized for $U \equiv X_1$, and thus we re-derive the capacity region of the less-noisy CRC without secrecy that is recently established in [12, Th. 10].

Now consider the Gaussian CRC (1). As mentioned earlier, if $ab = 1$ and also $|b| \le 1 < |a|$, the channel is degraded and satisfies the less-noisy condition in (4). In the next theorem, we explicitly derive the capacity-equivocation region for this case.

***Theorem 3)*** *Consider the Gaussian CRC (1) with confidential messages. If $ab = 1$ and also $|b| \le 1 < |a|$, the capacity-equivocation region is given by:*

$$\bigcup_{0 \le \alpha \le 1} \left\{ \begin{array}{l} (R_1,R_2,R_{e,1},R_{e,2}) \in \mathbb{R}_+^4 : \\ R_{e,1} \le R_1, \qquad R_{e,2} = 0 \\ R_1 \le \psi(\alpha P_1) \\ R_2 \le \psi\left(\dfrac{(1-\alpha)b^2P_1 + P_2 + 2|b|\sqrt{(1-\alpha)P_1P_2}}{\alpha b^2 P_1 + 1}\right) \\ R_{e,1} \le \psi(\alpha P_1) - \psi(\alpha b^2 P_1) \end{array} \right\} \tag{9}$$

*Proof of Theorem 3)* It is sufficient to show that the rate region (7) is optimized in consideration of Gaussian distributions. This can be proved using the entropy power inequality, as given in the Appendix. ∎

We next provide an achievable rate region for the CRC with confidential messages using the binning techniques. As we will see subsequently, this new rate region is useful to establish the capacity-equivocation region for some classes of discrete semi-deterministic channels and also Gaussian channels.

***Theorem 4)*** *Define the rate region $\mathfrak{R}_i$ as follows:*

$$\mathfrak{R}_i \triangleq$$

$$\bigcup_{P_{QWVUX_1X_2}} \left\{ \begin{array}{l} (R_1,R_2,R_{e,1},R_{e,2}) \in \mathbb{R}_+^4 : \\ R_{e,1} \le R_1, \qquad R_{e,2} \le R_2 \\ R_1 \le \min\left\{ \begin{array}{r} I(U;Y_1|Q) - I(U;W,X_2|Q), \\ I(U;Y_1|Q) + I(V;Y_2|W,X_2,Q) \\ -I(U;V,W,X_2|Q) \end{array} \right\} \\ R_2 \le I(V,W,X_2;Y_2|Q) \\ R_1 + R_2 \le I(U;Y_1|Q) + I(V,W,X_2;Y_2|Q) \\ \qquad\qquad -I(U;V,W,X_2|Q) \\ R_{e,1} \le [I(U;Y_1|Q) - I(U;Y_2,V,W,X_2|Q)]_+ \\ R_{e,2} \le \left[ \begin{array}{c} I(V;Y_2|W,X_2,Q) \\ -I(V;Y_1,U|W,X_2,Q) \end{array} \right]_+ \end{array} \right\} \tag{10}$$

*The set $\mathfrak{R}_i$ constitutes an inner bound on the capacity-equivocation region of the CRC in Fig. 1.*

*Proof of Theorem 4)* The detailed proof can be found in Appendix. Here, we present the key ingredients of our achievability scheme. First note that since the primary transmitter applies a deterministic encoder, it is seldom effective to achieve secrecy for its message. To compensate for this, in our coding scheme the primary transmitter splits its messages into two parts: $M_2 = (M_{21}, M_{22})$, and then ignores the sub-message $M_{21}$ and relegates its transmission to the cognitive transmitter. Thus, the cognitive transmitter contributes to achieve secrecy for the primary transmitter's message in addition to its own message. It is worth stating that for the CRC without secrecy, relegating transmission of a part of primary transmitter's message to the cognitive transmitter is not helpful in improving the resultant achievable rate region, as shown in [13, Sec. III.A.4]. Nevertheless, for the channel with confidential messages the latter technique is beneficial to achieve a positive secrecy rate for the primary transmitter's message.

The messages $M_1$, $M_{21}$ and $M_{22}$ are next encoded by a random code of length-$n$. For simplicity of exposition, let assume that $W \equiv Q \equiv \emptyset$. The primary transmitter encodes the sub-message $M_{22}$ by a codeword $X_2^n$ generated based on $P_{X_2}(x_2)$, and sends it over the channel. Encoding at the cognitive transmitter includes a random binning scheme. The binning that we apply is different from the double binning used in [18] and [19] to derive achievable rate-equivocation region for the two-user broadcast channel. The double binning technique of [18] and [19] combines two phases of binning: one for joint precoding as in the Marton's scheme for the broadcast channel without secrecy [24] and the other for preserving of confidentiality of messages. This technique requires a complicated scheme for codeword assigning to messages (see for example [19, page 4537]). In our coding scheme only a single phase of binning is required. At the cognitive transmitter a bin of random codewords $U^n$ conveying the message $M_1$, and also a bin of codewords $V^n$ conveying the sub-message $M_{21}$ are generated. The codewords $U^n$ are generated based on $P_U(u)$, and the codewords $V^n$ are superimposed on the codeword $X_2^n$ and generated based on $P_{V|X_2}(v|x_2)$. The sizes of the bins are intelligently designed to guarantee that: 1) The set of all jointly typical triples $(X_2^n, V^n, U^n)$ is nonempty, 2) The confidentiality of messages is preserved. Thus, we no longer need to the double binning. The benefit of our approach is that it does not involve a complicated scheme for codeword assigning to messages. Superimposed on a jointly typical triple $(X_2^n, V^n, U^n)$ that is randomly and uniformly chosen from all such (jointly typical) triples, the cognitive transmitter then generates a codeword $X_1^n$ based on $P_{X_1|UVX_2}(x_1|u, v, x_2)$ and sends it over the channel. For decoding, each receiver applies a jointly typical decoder. The receiver $Y_1$ explores within the codewords $U^n$ to detect the message $M_1$, and the receiver $Y_2$ within the codewords $X_2^n, V^n$ to detect the messages $M_{22}$ and $M_{21}$. Analysis of this coding scheme leads to the desired rate region. ■

We now show that the achievable rate region (10) coincides with the outer bound (2) for some special cases. Consider a class of CRCs for which the following inequality holds:

$$H(Y_2|W) - H(Y_2|X_2) \leq H(Y_1|W) - H(Y_1|X_2) \tag{11}$$

for all joint PDFs $P_{WX_1X_2}(w, x_1, x_2)$.

In the next theorem, we establish the capacity-equivocation region of the semi-deterministic CRC (the received signal at the cognitive receiver is given by a deterministic function of the channel inputs) which also satisfies the condition (11).

***Theorem 5)*** *Consider the CRC with confidential messages as shown in Fig. 1. Assume that the channel is semi-deterministic in the sense of $Y_1 = f(X_1, X_2)$ where $f(.)$ is an arbitrary deterministic function. If the channel also satisfies the condition (11), then the capacity-equivocation region is given by:*

$$\bigcup_{P_{VX_1X_2}} \left\{ \begin{array}{l} (R_1, R_2, R_{e,1}, R_{e,2}) \in \mathbb{R}_+^4 : \\ R_{e,1} \leq R_1, \qquad R_{e,2} \leq R_2 \\ R_1 \leq \min\left\{ \begin{array}{c} H(Y_1|X_2), \\ H(Y_1|V, X_2) + I(V; Y_2|X_2) \end{array} \right\} \\ R_2 \leq I(V, X_2; Y_2) \\ R_1 + R_2 \leq H(Y_1|V, X_2) + I(V, X_2; Y_2) \\ R_{e,1} \leq H(Y_1|Y_2, V, X_2) \\ R_{e,2} \leq [I(V; Y_2|X_2) - I(V; Y_1|X_2)]_+ \end{array} \right\} \tag{12}$$

*Proof of Theorem 5)* The achievability is derived by setting $U \equiv Y_1$ and $W \equiv Q \equiv \emptyset$ in $\mathfrak{R}_i$ given by (10). For the converse part consider the outer bound $\mathfrak{R}_o^{UVW}$ in (2); we have:

$$R_1 \leq I(U, V; Y_1|X_2) \leq H(Y_1|X_2) \tag{13}$$

$$\begin{aligned} R_1 &\leq I(U; Y_1|V, X_2) + I(V; Y_2|X_2) \\ &\leq H(Y_1|V, X_2) + I(V; Y_2|X_2) \end{aligned} \tag{14}$$

$$R_2 \leq I(V, X_2; Y_2) \tag{15}$$

$$\begin{aligned} R_1 + R_2 &\leq I(U; Y_1|V, X_2) + I(V, X_2; Y_2) \\ &\leq H(Y_1|V, X_2) + I(V, X_2; Y_2) \end{aligned} \tag{16}$$

$$\begin{aligned} R_{e,1} &\leq [I(U; Y_1|V, X_2) - I(U; Y_2|V, X_2)]_+ \\ &\leq [I(U; Y_1, Y_2|V, X_2) - I(U; Y_2|V, X_2)]_+ \\ &= I(U; Y_1|Y_2, V, X_2) \leq H(Y_1|Y_2, V, X_2) \end{aligned} \tag{17}$$

$$\begin{aligned} R_{e,2} &\leq [I(V, X_2; Y_2|W) - I(V, X_2; Y_1|W)]_+ \\ &\overset{(a)}{=} \left[ \begin{array}{c} (H(Y_2|W) - H(Y_1|W)) \\ +(H(Y_1|V, X_2) - H(Y_2|V, X_2)) \end{array} \right]_+ \\ &\overset{(b)}{\leq} \left[ \begin{array}{c} (H(Y_2|X_2) - H(Y_1|X_2)) \\ +(H(Y_1|V, X_2) - H(Y_2|V, X_2)) \end{array} \right]_+ \\ &= [I(V; Y_2|X_2) - I(V; Y_1|X_2)]_+ \end{aligned} \tag{18}$$

where equality (a) holds because $W \to V, X_2 \to Y_1, Y_2$ forms a Markov chain (see the distribution of the outer bound $\mathfrak{R}_o^{UVW}$ in (2)); also, inequality (b) is due to (11). The proof is thus complete. ■

***Remarks 2:***

1. Consider the condition (11). By substituting $W \equiv (V, X_2)$, it is reduced to:

$$I(V; Y_1|X_2) \leq I(V; Y_2|X_2)$$

   for all joint PDFs $P_{VX_1X_2}(v, x_1, x_2)$. Therefore, for the channels satisfying (11) positive equivocation rate can be achieved for the message of the primary transmitter, i.e., $M_2$.

2. Consider the following constraint from the outer bound $\mathfrak{R}_o^{UVW}$ given by (2):

$$R_1 \leq I(U; Y_1|V, X_2) + I(V; Y_2|X_2)$$

   In such constraints a single communication rate, i.e., $R_1$, is bounded above by the addition of two mutual information functions. Such constraints are not given in any of the outer bounds derived in previous papers [9-12, 22] for the CRC. Theorem 5 demonstrates that these constraints can be useful to establish capacity results for channels with confidential messages. This theorem also shows that those constraints of the outer bound $\mathfrak{R}_o^{UVW}$ in (2) which include the auxiliary random variable $W$ are helpful to prove capacity results.

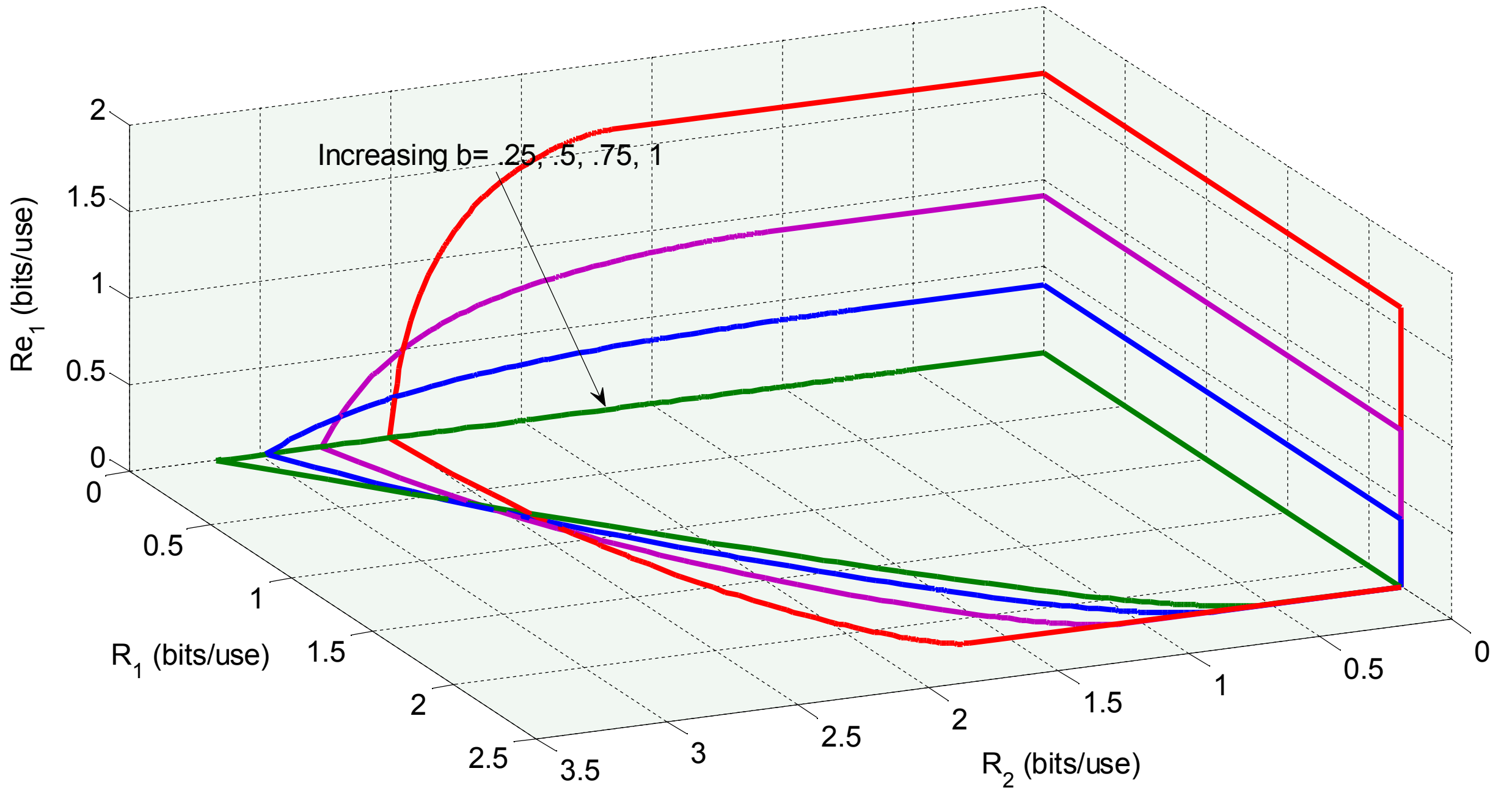

Figure 2. The capacity-equivocation region of the Gaussian CRC in (1) where the cognitive message $M_1$ is confidential but there is no secrecy constraint for the primary message $M_2$. The power constraints are given by $P_1 = P_2 = 20$ and the parameter $a$ is equal to 1.

In all cases considered above, it was assumed that both messages are required to be kept confidential from the non-corresponding receiver. In other words, we were interested in achieving secrecy for both messages. Since we have imposed that the primary transmitter uses a deterministic encoder, this transmitter is seldom effective in increasing the secrecy for its message. Therefore, achieving secrecy for the message of the primary transmitter ($M_2$ in Fig. 1) to some extent is more difficult than that for the message of the cognitive transmitter ($M_1$ in Fig. 1). In the sequel, we relax the secrecy constraint of the primary transmitter's message and study a case where only the cognitive transmitter's message is required to be kept confidential. This case was also studied in [23] where only a simple achievable rate region is given. Here, we derive some important capacity results for the case, specifically, for the semi-deterministic and Gaussian channels. The results are given in the following theorems.

***Theorem 6)*** *Consider the CRC in Fig. 1 where the message $M_1$ is confidential but there is no confidentiality requirement for the message $M_2$. The capacity-equivocation region of the semi-deterministic channel with $Y_1 = f(X_1, X_2)$, where $f(.)$ is an arbitrary deterministic function, is given by:*

$$\bigcup_{P_{VX_1X_2}} \left\{ \begin{array}{l} (R_1, R_2, R_{e,1}) \in \mathbb{R}_+^3: \\ R_{e,1} \leq R_1, \\ R_1 \leq \min\left\{ \begin{array}{c} H(Y_1|X_2), \\ H(Y_1|V,X_2) + I(V;Y_2|X_2) \end{array} \right\} \\ R_2 \leq I(V,X_2;Y_2) \\ R_1 + R_2 \leq H(Y_1|V,X_2) + I(V,X_2;Y_2) \\ R_{e,1} \leq H(Y_1|Y_2,V,X_2) \end{array} \right\} \tag{19}$$

*Proof of Theorem 6)* The achievability is obtained by setting $U \equiv Y_1$, $W \equiv Q \equiv \emptyset$, and $R_{e,2} = 0$ in $\mathfrak{R}_i$ given by (10). The converse part is derived from the outer bound $\mathfrak{R}_o^{UVW}$ in (2) exactly similar to the converse part of Theorem 5. Note that, in Theorem 5, the condition (11) was only used when bounding the equivocation rate $R_{e,2}$, as given in (18); this step is out of interest in Theorem 6. ■

***Remark 3:*** By setting $R_{e,1} = 0$ in (19), we re-derive the capacity region of the semi-deterministic CRC without secrecy which is recently obtained in [12, Th. 11].

***Theorem 7)*** *Consider the CRC in Fig. 1 where the message $M_1$ is confidential but there is no confidentiality requirement for the message $M_2$. The capacity-equivocation region of the Gaussian channel in (1) with $|b| \leq 1$ is given by:*

$$\bigcup_{0 \leq \alpha \leq 1} \left\{ \begin{array}{l} (R_1, R_2, R_{e,1}) \in \mathbb{R}_+^3: \\ R_{e,1} \leq R_1, \\ R_1 \leq \psi(\alpha P_1) \\ R_2 \leq \psi\left( \dfrac{(1-\alpha)b^2 P_1 + P_2 + 2|b|\sqrt{(1-\alpha)P_1 P_2}}{\alpha b^2 P_1 + 1} \right) \\ R_{e,1} \leq \psi(\alpha P_1) - \psi(\alpha b^2 P_1) \end{array} \right\} \tag{20}$$

*Proof of Theorem 7)* The direct part is derived by setting $V \equiv Q \equiv \emptyset$ and $R_{e,2} = 0$ in the achievable rate region (10). Also, the converse part is proved using the outer bound (2). The complete proof is given in the Appendix. ■

We remark that the coding strategy that achieves the capacity in Theorem 7 is actually the dirty paper coding scheme [26]. In this coding scheme, the signal conveying the primary message is treated as (known) side information at the cognitive transmitter and its interference effect on the cognitive receiver is canceled (exactly similar to the interference cancellation in the dirty paper channel [26]). As a result, the capacity-equivocation region given in (20) does not depend on the channel parameter $a$ because the interference effect of the term $aX_2$ on the receiver $Y_1$ is completely canceled.

Figure 2 on the top of the previous page depicts the capacity-equivocation region of the Gaussian CRC in (1) where the cognitive message $M_1$ is confidential but there is no secrecy constraint for the primary message $M_2$, as given by Theorem 7. This figure is due to $P_1 = P_2 = 20$, $a = 1$, and $b = .25, .5, .75$, and $1$. As shown in the figure, when $b$ increases, the maximum achievable rate for $R_2$ is improved because the signal to noise ratio at the receiver $Y_2$ increases; also, the receiver $Y_2$ can obtain more information about the non-respective message $M_1$, and therefore the equivocation rate $R_{e,1}$ decreases. When $b = 1$, the receiver $Y_2$ can fully decode the message $M_1$ and hence it is not possible to transmit $M_1$ with a positive rate securely, i.e., $R_{e,1} = 0$.

Based on Corollaries 1 and 2 and Theorem 7, we can directly derive a full characterization of the capacity region of the *general* Gaussian CRC with perfect secrecy for the message of the cognitive transmitter but no secrecy constraint for the message of the primary transmitter. This result is given in the following corollary.

***Corollary 3)*** *Consider the CRC in Fig. 1 where the message $M_1$ is required to be transmitted with perfect secrecy but there is no secrecy constraint for the message $M_2$. The secrecy capacity region of the general Gaussian channel (1) is given below:*

$$\bigcup_{0\le\alpha\le1}\left\{\begin{array}{l}(R_1,R_2)\in\mathbb{R}_+^2:\\ R_1\le[\psi(\alpha P_1)-\psi(\alpha b^2P_1)]_+\\ R_2\le\psi\left(\dfrac{(1-\alpha)b^2P_1+P_2+2|b|\sqrt{(1-\alpha)P_1P_2}}{\alpha b^2P_1+1}\right)\end{array}\right\} \tag{21}$$

*Proof of Corollary 3)* For the case of $|b|\le1$, the secrecy capacity region is derived from Theorem 7 by setting $R_1 = R_{e,1}$ in (20). For the case of $|b|>1$, according to Corollaries 1 and 2, no secrecy can be achieved for the message $M_1$; thereby, $R_1 = R_{e,1} = 0$. Note that the maximum achievable rate for the message $M_2$ is given by $\max_{P_{X_1X_2}} I(X_1,X_2;Y_2)$. Thus, the secrecy capacity region is as follows:

$$\bigcup\left\{\begin{array}{l}(R_1,R_2)\in\mathbb{R}_+^2:\\ R_1=0\\ R_2\le\psi\left(b^2P_1+P_2+2|b|\sqrt{P_1P_2}\right)\end{array}\right\} \tag{22}$$

Note that when $|b|>1$, the characterization (21) is equivalent to (22). ■

It is interesting that unlike the Gaussian CRC without secrecy for which the capacity region is in general unknown, by imposing the perfect secrecy constraint for the cognitive message we can fully characterize the capacity, as given in Corollary 3.

## Conclusion

In this paper, we explored fundamental capacity limits for the cognitive radio channel with confidential messages. We established inner and outer bounds for the capacity-equivocation region. Compared to the outer bounds previously given in [9-12, 22] for the CRC (without secrecy), ours has a novel structure that makes it more efficient to derive capacity results. Also, our inner bound was derived by a novel approach that requires a simpler scheme for codeword assigning to messages than the double binning technique of [18] and [19]. We then showed that the obtained inner and outer bounds yield exact capacity for some special cases. Specifically, we derived the capacity-equivocation region for a class of less-noisy CRCs and also a class of semi-deterministic CRCs. For the case where only the message of the cognitive transmitter is required to be kept confidential, we also established the capacity-equivocation region for the Gaussian CRC with weak interference. Thus, in this paper the first capacity results for the CRC with confidential messages were established.

Here, we studied the case where the cognitive transmitter applies a stochastic encoder and the primary transmitter applies a deterministic one. An interesting next step would be to analyze for the case where both transmitters apply stochastic encoders.

## Appendix

***Proof of Theorem 1)*** To obtain $\mathfrak{R}_o^{UVW}$ in (2), we first adapt the outer bound derived for the CRC in [13, Sec. III.B] to include secrecy constraints, as given in the following lemma.

*Lemma 1) Define the rate region $\mathfrak{R}_o$ as follows:*

$$\mathfrak{R}_o \triangleq$$
$$\bigcup_{\mathcal{P}_o}\left\{\begin{array}{l}(R_1,R_2,R_{e,1},R_{e,2})\in\mathbb{R}_+^4:\\ R_{e,1}\le R_1,\qquad R_{e,2}\le R_2,\\ R_1\le\min\left\{\begin{array}{c}I(Z,M_1;Y_1|Q),I(Z,M_1;Y_1|M_2,Q),\\ I(M_1;Y_1|Z,Q)+I(Z;Y_2|Q),\\ I(M_1;Y_1|Z,M_2,Q)+I(Z;Y_2|M_2,Q)\end{array}\right\}\\ R_2\le\min\left\{\begin{array}{c}I(Z,M_2;Y_2|Q),I(Z,M_2;Y_2|M_1,Q),\\ I(M_2;Y_2|Z,Q)+I(Z;Y_1|Q),\\ I(M_2;Y_2|Z,M_1,Q)+I(Z;Y_1|M_1,Q)\end{array}\right\}\\ R_1+R_2\le I(M_1;Y_1|Z,M_2,Q)+I(Z,M_2;Y_2|Q)\\ R_1+R_2\le I(M_2;Y_2|Z,M_1,Q)+I(Z,M_1;Y_1|Q)\\ R_1+R_2\le I(M_1;Y_1|Z,M_2,Q)+I(M_2;Y_2|Z,Q)\\ \qquad\qquad +I(Z;Y_1|Q)\\ R_1+R_2\le I(M_2;Y_2|Z,M_1,Q)+I(M_1;Y_1|Z,Q)\\ \qquad\qquad +I(Z;Y_2|Q)\\ R_{e,1}\le[I(M_1;Y_1|Z,Q)-I(M_1;Y_2|Z,Q)]_+\\ R_{e,1}\le[I(M_1;Y_1|Z,M_2,Q)-I(M_1;Y_2|Z,M_2,Q)]_+\\ R_{e,2}\le[I(M_2;Y_2|Z,Q)-I(M_2;Y_1|Z,Q)]_+\\ R_{e,2}\le[I(M_2;Y_2|Z,M_1,Q)-I(M_2;Y_1|Z,M_1,Q)]_+\end{array}\right\} \tag{A~1}$$

*where $\mathcal{P}_o$ denotes the set of all joint PDFs $P_{QM_1M_2ZX_1X_2}(q,m_1,m_2,z,x_1,x_2)$ satisfying:*

$$P_{QM_1M_2ZX_1X_2}=P_QP_{M_1}P_{M_2}P_{X_1|M_1M_2Q}P_{X_2|M_2Q}P_{Z|X_1X_2M_1M_2Q} \tag{A~2}$$

*Also, the input signal $X_2$ is given by a deterministic function of $(M_2,Q)$; in other words, we have $P_{X_2|M_2Q}\in\{0,1\}$. The rate region $\mathfrak{R}_o$ constitutes an outer bound for the capacity-equivocation region of the CRC in Fig. 1.*

*Proof of Lemma 1)* Consider a code of sufficiently large length $n$ with vanishing error probability for the channel. First note that:

$$\begin{cases}R_{e,1}\le\frac{1}{n}H(M_1|Y_2^n)\le\frac{1}{n}H(M_1)=R_1\\ R_{e,2}\le\frac{1}{n}H(M_2|Y_1^n)\le\frac{1}{n}H(M_2)=R_2\end{cases}$$

Also, using Fano's inequality we have:

$$\frac{1}{n}H(M_i|Y_i^n) \leq \epsilon_{i,n}, \; i = 1,2$$

where $\epsilon_{1,n} \to 0$ and $\epsilon_{2,n} \to 0$ as $n \to \infty$. Define the random variables $Z_t, t = 1, \dots, n$, as $Z_t \triangleq (Y_1^{t-1}, Y_{2,t+1}^n)$. The constraints on the communication rates $R_1$ and $R_2$ can be established exactly similar to the ones derived in [13, Sec. III.B]. Here, we do not repeat the derivations for brevity. Therefore, it is only required to prove the constraints on the equivocation rates $R_{e,1}$ and $R_{e,2}$. To this end, we can write:

$$\begin{aligned} nR_{e,1} &\leq H(M_1|Y_2^n) \\ &= H(M_1|Y_2^n, M_2) + I(M_1; M_2|Y_2^n) \\ &= H(M_1|M_2) - I(M_1; Y_2^n|M_2) + I(M_1; M_2|Y_2^n) \\ &= I(M_1; Y_1^n|M_2) - I(M_1; Y_2^n|M_2) + H(M_1|Y_1^n, M_2) \\ &\qquad + I(M_1; M_2|Y_2^n) \\ &\leq I(M_1; Y_1^n|M_2) - I(M_1; Y_2^n|M_2) + n(\epsilon_{1,n} + \epsilon_{2,n}) \end{aligned} \quad \text{(A~3)}$$

For the first term of (A~3) we have:

$$\begin{aligned} I(M_1; Y_1^n|M_2) &= \textstyle\sum_{t=1}^n I(M_1; Y_{1,t}|Y_1^{t-1}, M_2) \\ &= \textstyle\sum_{t=1}^n I(M_1, Y_{2,t+1}^n; Y_{1,t}|Y_1^{t-1}, M_2) \\ &\qquad - \textstyle\sum_{t=1}^n I(Y_{2,t+1}^n; Y_{1,t}|Y_1^{t-1}, M_1, M_2) \\ &= \textstyle\sum_{t=1}^n I(M_1; Y_{1,t}|Y_1^{t-1}, Y_{2,t+1}^n, M_2) \\ &\qquad + \textstyle\sum_{t=1}^n I(Y_{2,t+1}^n; Y_{1,t}|Y_1^{t-1}, M_2) \\ &\qquad - \textstyle\sum_{t=1}^n I(Y_{2,t+1}^n; Y_{1,t}|Y_1^{t-1}, M_1, M_2) \\ &= \textstyle\sum_{t=1}^n I(M_1; Y_{1,t}|Y_1^{t-1}, Y_{2,t+1}^n, M_2) \\ &\qquad + \textstyle\sum_{t=1}^n I(Y_1^{t-1}; Y_{2,t}|Y_{2,t+1}^n, M_2) \\ &\qquad - \textstyle\sum_{t=1}^n I(Y_1^{t-1}; Y_{2,t}|Y_{2,t+1}^n, M_1, M_2) \end{aligned} \quad \text{(A~4)}$$

where the last equality of (A~4) is due to Csiszar-Korner identity [16]. Also, for the second term of (A~3) we have:

$$\begin{aligned} I(M_1; Y_2^n|M_2) &= \textstyle\sum_{t=1}^n I(M_1; Y_{2,t}|Y_{2,t+1}^n, M_2) \\ &= \textstyle\sum_{t=1}^n I(M_1, Y_1^{t-1}; Y_{2,t}|Y_{2,t+1}^n, M_2) \\ &\qquad - \textstyle\sum_{t=1}^n I(Y_1^{t-1}; Y_{2,t}|Y_{2,t+1}^n, M_1, M_2) \\ &= \textstyle\sum_{t=1}^n I(M_1; Y_{2,t}|Y_1^{t-1}, Y_{2,t+1}^n, M_2) \\ &\qquad + \textstyle\sum_{t=1}^n I(Y_1^{t-1}; Y_{2,t}|Y_{2,t+1}^n, M_2) \\ &\qquad - \textstyle\sum_{t=1}^n I(Y_1^{t-1}; Y_{2,t}|Y_{2,t+1}^n, M_1, M_2) \end{aligned} \quad \text{(A~5)}$$

By substituting (A~4) and (A~5) in (A~3), we derive:

$$\begin{aligned} nR_{e,1} &\leq \textstyle\sum_{t=1}^n I(M_1; Y_{1,t}|Z_t, M_2) - \textstyle\sum_{t=1}^n I(M_1; Y_{2,t}|Z_t, M_2) \\ &\qquad + n(\epsilon_{1,n} + \epsilon_{2,n}) \end{aligned} \quad \text{(A~6)}$$

One can show that all the equations (A~3)-(A~6) are still valid if the message $M_2$ is removed everywhere. Thus, we derive:

$$\begin{aligned} nR_{e,1} &\leq \textstyle\sum_{t=1}^n I(M_1; Y_{1,t}|Z_t) \\ &\qquad - \textstyle\sum_{t=1}^n I(M_1; Y_{2,t}|Z_t) + n(\epsilon_{1,n} + \epsilon_{2,n}) \end{aligned} \quad \text{(A~7)}$$

Also, by following a rather similar procedure, one can obtain:

$$\begin{aligned} nR_{e,2} &\leq \textstyle\sum_{t=1}^n I(M_2; Y_{2,t}|Z_t, M_1) - \textstyle\sum_{t=1}^n I(M_2; Y_{1,t}|Z_t, M_1) \\ &\qquad + n(\epsilon_{1,n} + \epsilon_{2,n}) \\ nR_{e,2} &\leq \textstyle\sum_{t=1}^n I(M_2; Y_{2,t}|Z_t) - \textstyle\sum_{t=1}^n I(M_2; Y_{1,t}|Z_t) \\ &\qquad + n(\epsilon_{1,n} + \epsilon_{2,n}) \end{aligned} \quad \text{(A~8)}$$

Now let $Q$ be a time-sharing variable uniformly distributed over the set $\{1, \dots, n\}$. Also, define:

$$Z \triangleq Z_Q, \qquad Y_i \triangleq Y_{i,Q}, \qquad X_i \triangleq X_{i,Q}, \qquad i = 1,2$$

then by applying this time-sharing to inequalities (A~6)-(A~8) and letting $n$ tend to infinity, we obtain the desired constraints on the equivocation rates as in (2). Note that for the channel with secrecy (unlike the case of without secrecy) due to the stochastic encoding at the cognitive transmitter, the following Markov chain does not necessarily hold (for any $t \in \{1, \dots, n\}$):

$$X_{1,t}, X_{2,t} \to M_1, M_2 \to Z_t$$

This justifies the distribution (A~2). Also, since the primary transmitter applies a deterministic encoder, for every $t = 1, \dots, n$, the signal $X_{2,t}$ is given by a deterministic function of the message $M_2$. Therefore, the input variable $X_2$ (that is actually $X_{2,Q}$) is also a deterministic function of $(M_2, Q)$. ■

It is worth noting that the outer bound (A~1) is given for the CRC shown in Fig. 1, however, the same bound could be derived for the two-user CIC and the two-user BC with confidential messages. For each case, it is only required to modify the joint PDFs (A~2) according to the corresponding network topology.

Now using $\mathfrak{R}_o$ given in (A~1) we derive the desired outer bound, i.e., $\mathfrak{R}_o^{UVW}$ in (2). Define:

$$U \triangleq (Z, M_1, Q), \qquad V \triangleq (Z, M_2, Q), \qquad W \triangleq (Z, Q)$$

Note that since the primary transmitter applies a deterministic encoder, $X_2$ is a deterministic function of $(M_2, Q)$. Moreover, $W \to U \to X_1, X_2$ and $W \to V \to X_1, X_2$ form Markov chains because $W$ is actually a part of $U$ and $V$. Now, we can write:

$$\begin{aligned} R_1 \leq I(Z, M_1; Y_1|M_2, Q) &= I(Z, M_1; Y_1|M_2, Q, X_2) \\ &\leq I(Z, M_1, M_2, Q; Y_1|X_2) \\ &= I(U, V; Y_1|X_2) \end{aligned}$$

The other constraints of (2) can also be derived based on the outer bound $\mathfrak{R}_o$ in (A~1) similarly. The proof of Theorem 1 is thus complete. We remark that the outer bound $\mathfrak{R}_o$ in (A~1) is (potentially) tighter than $\mathfrak{R}_o^{UVW}$ in (2) but its evaluation is more complex. Also, the characterization of $\mathfrak{R}_o^{UVW}$ enables us to prove capacity results more flexibly. ■

***Proof of Theorem 3)*** Consider the rate region (7). We prove that for the Gaussian channel (1) with $ab = 1$ and $|b| \leq 1 < |a|$, it is reduced to (9). Let $X_2$ be a Gaussian distributed random variable with zero mean and variance $P_2$. Also, let $\tilde{X}_1$

be a Gaussian random variable independent of $X_2$ with zero mean and unit variance. Define:

$$\begin{cases} X_1 \triangleq \frac{|b|}{b}\sqrt{(1-\alpha)\frac{P_1}{P_2}}X_2 + \sqrt{\alpha P_1}\tilde{X}_1, & 0 \le \alpha \le 1 \\ U \equiv X_1 \\ V \equiv \emptyset \end{cases}$$

By substituting $X_1, X_2, U$, and $V$ in (7), we obtain the achievability of (9). Now we prove that the Gaussian inputs are actually optimal. First note that since $|b| \le 1$, one can show that:

$$\begin{cases} I(U;Y_2|V,X_2) \le I(U;Y_1|V,X_2) \\ I(X_1;Y_2|U,V,X_2) \le I(X_1;Y_1|U,V,X_2) \end{cases} \tag{A~9}$$

for all joint PDFs $P_{UVX_1X_2}$. In fact, since $|b| \le 1$, given $X_2$, the output $Y_2$ is a stochastically degraded version of $Y_1$ and therefore both of inequalities in (A~9) hold. Now for the mutual information functions given in rate region (7), we can write:

$$\begin{aligned}
&I(U,V;Y_1|X_2) \le I(X_1,U,V;Y_1|X_2) = I(X_1;Y_1|X_2) \\
&I(U;Y_1|V,X_2) \le I(X_1,U;Y_1|V,X_2) = I(X_1;Y_1|V,X_2) \\
&[I(U;Y_1|V,X_2) - I(U;Y_2|V,X_2)]_+ \\
&\quad = \begin{bmatrix} \big(I(X_1,U;Y_1|V,X_2) - I(X_1,U;Y_2|V,X_2)\big) \\ -\big(I(X_1;Y_1|U,V,X_2) - I(X_1;Y_2|U,V,X_2)\big) \end{bmatrix}_+ \\
&\quad \overset{(a)}{\le} [I(X_1,U;Y_1|V,X_2) - I(X_1,U;Y_2|V,X_2)]_+ \\
&\quad = [I(X_1;Y_1|V,X_2) - I(X_1;Y_2|V,X_2)]_+
\end{aligned} \tag{A~10}$$

where inequality (a) is due to (A~9). Considering equations (A~10), we deduce that the rate region (7) is optimized for $U \equiv X_1$. By this choice, we obtain the following rate region:

$$\bigcup_{P_{VX_1X_2}} \left\{ \begin{aligned} &(R_1,R_2,R_{e,1},R_{e,2}) \in \mathbb{R}_+^4: \\ &R_{e,1} \le R_1, \qquad R_{e,2} = 0 \\ &R_1 \le I(X_1;Y_1|X_2) \\ &R_2 \le I(V,X_2;Y_2) \\ &R_1 + R_2 \le I(X_1;Y_1|V,X_2) + I(V,X_2;Y_2) \\ &R_{e,1} \le [I(X_1;Y_1|V,X_2) - I(X_1;Y_2|V,X_2)]_+ \end{aligned} \right\} \tag{A~11}$$

Note that since $|b| \le 1$, given $X_2$, the output $Y_2$ is a (stochastically) degraded version of $Y_1$, and we thus have:

$$I(V;Y_2|X_2) \le I(V;Y_1|X_2) \quad \text{for all joint PDFs } P_{VX_1X_2}$$

Now consider the following rate region:

$$\bigcup_{P_{VX_1X_2}} \left\{ \begin{aligned} &(R_1,R_2,R_{e,1},R_{e,2}) \in \mathbb{R}_+^4: \\ &R_{e,1} \le R_1, \qquad R_{e,2} = 0 \\ &R_1 \le I(X_1;Y_1|V,X_2) \\ &R_2 \le I(V,X_2;Y_2) \\ &R_{e,1} \le [I(X_1;Y_1|V,X_2) - I(X_1;Y_2|V,X_2)]_+ \end{aligned} \right\} \tag{A~12}$$

This region is a convex set (see the derivation of the outer bound (2) in Appendix where the time-sharing parameter $Q$ is absorbed into the auxiliary $V$). It is clear that (A~12) is a subset of (A~11). On the one hand, by following the same arguments as [9, Prop. 3.3], one can prove that every point of the region (A~11) belongs to the convex closure of (A~12). Since (A~12) is convex, we deduce that (A~11) is a subset of (A~12). Thus, the regions (A~11) and (A~12) are equivalent. We next show that the Gaussian distributions optimize the rate region (A~12). First, let us remark that the Gaussian CRC (1) with $|b| \le 1$ is equivalent to the following:

$$\begin{cases} Y_1 = X_1 + aX_2 + Z_1 \\ \tilde{Y}_2 = bY_1 + (1-ab)X_2 + \sqrt{1-b^2}\tilde{Z}_2 \end{cases},$$

where $\tilde{Z}_2$ is a Gaussian RV (independent of $Z_1$) with zero mean and unit variance. Define:

$$\rho \triangleq \frac{\mathbb{E}[X_1X_2]}{\sqrt{P_1P_2}}$$

We have:

$$\begin{aligned} H(Y_1|V,X_2) &\le H(X_1+Z_1|V,X_2) \\ &\le H(X_1+Z_1|X_2) \\ &\overset{(a)}{\le} \tfrac{1}{2}\log 2\pi e\big((1-\rho^2)P_1 + 1\big) \end{aligned} \tag{A~13}$$

Also,

$$H(Y_1|V,X_2) \ge H(Y_1|X_1,V,X_2) = H(Z_1) = \frac{1}{2}\log 2\pi e \tag{A~14}$$

where inequality (a) is due to the "Gaussian maximizes entropy" principle. Considering (A~13) and (A~14), we deduce that there exists $\alpha \in [0,1]$ so that:

$$H(Y_1|V,X_2) = \tfrac{1}{2}\log 2\pi e(\alpha P_1 + 1), \quad \text{and} \quad |\rho| \le \sqrt{1-\alpha} \tag{A~15}$$

Then, let evaluate $H(Y_2|V,X_2)$. We have:

$$\begin{aligned} H(Y_2|V,X_2) &= H(\tilde{Y}_2|V,X_2) = H\left(bY_1 + \sqrt{1-b^2}\tilde{Z}_2\middle|V,X_2\right) \\ &\overset{(a)}{\ge} \frac{1}{2}\log\left(2^{2H(bY_1|V,X_2)} + 2^{2H\left(\sqrt{1-b^2}\tilde{Z}_2|V,X_2\right)}\right) \\ &\overset{(b)}{=} \frac{1}{2}\log\left(2\pi e\big(b^2(\alpha P_1+1)\big) + 2\pi e(1-b^2)\right) \\ &= \frac{1}{2}\log\big(2\pi e(\alpha b^2P_1 + 1)\big) \end{aligned} \tag{A~16}$$

where (a) is due to the entropy power inequality [27], and (b) is due to (A~15). Now, we can write:

$$I(X_1;Y_1|V,X_2) \overset{(a)}{=} \psi(\alpha P_1) \tag{A~17}$$

where (a) is due to (A~15). Also,

$$\begin{aligned} I(V,X_2;Y_2) &= H(Y_2) - H(Y_2|V,X_2) \\ &\overset{(a)}{\le} H(bX_1 + X_2 + Z_2) - \tfrac{1}{2}\log\big(2\pi e(\alpha b^2P_1 + 1)\big) \end{aligned}$$

$$\overset{(b)}{\leq} \frac{1}{2}\log\left(2\pi e\left(b^2P_1 + P_2 + 2b\rho\sqrt{P_1P_2} + 1\right)\right)$$
$$-\frac{1}{2}\log\left(2\pi e(\alpha b^2 P_1 + 1)\right)$$
$$\overset{(c)}{\leq} \frac{1}{2}\log\left(2\pi e\left(b^2P_1 + P_2 + 2|b|\sqrt{(1-\alpha)P_1P_2} + 1\right)\right)$$
$$-\frac{1}{2}\log\left(2\pi e(\alpha b^2 P_1 + 1)\right)$$
$$= \psi\left(\frac{(1-\alpha)b^2P_1 + P_2 + 2|b|\sqrt{(1-\alpha)P_1P_2}}{\alpha b^2 P_1 + 1}\right)$$
(A~18)

where (a) is due to (A~16), (b) is due to the "Gaussian maximizes entropy" principle, and (c) holds because $|\rho| \leq \sqrt{1-\alpha}$ (see (A~15)). Finally, we have:

$$I(X_1;Y_1|V,X_2) - I(X_1;Y_2|V,X_2) \overset{(a)}{\leq} \psi(\alpha P_1) - \psi(\alpha b^2 P_1)$$
(A~19)

where (a) is due to (A~15) and (A~16). By substituting (A~17)-(A~19) in (A~12), we get the region (9), as desired. ■

***Proof of Theorem 4)*** Consider the CRC in Fig. 1 with confidential messages. We apply a random coding strategy to obtain the achievable rate region $\mathfrak{R}_i$ in (10). Note that the parameter $Q$ in (10) is a time-sharing random variable. We prove the achievability for the case of $Q \equiv \emptyset$. The coding strategy readily extends to include the time-sharing scheme. Also, for simplicity of exposition without loss of generality we assume that $W \equiv \emptyset$. To derive the desired rate region in (10), it is sufficient to re-define $X_2$ with $(W, X_2)$ in the coding scheme described below.

Fix a joint PDF $P_{VUX_1X_2}$ and its marginal distributions $P_{X_2}, P_{V|X_2}, P_U$ and $P_{X_1|UVX_2}$. Let $(R_1, R_2, R_{e,1}, R_{e,2})$ be a 4-tuple belonging to $\mathfrak{R}_i$. Without loss of generality assume that:

$$\begin{aligned} 0 \leq R_{e,1} &\leq I(U;Y_1) - I(U;Y_2,V,X_2) \\ 0 \leq R_{e,2} &\leq I(V;Y_2|X_2) - I(V;Y_1,U|X_2) \end{aligned}$$
(A~20)

Split the message $M_2$ and its respective rate $R_2$ into two parts as follows:

$$M_2 = (M_{21}, M_{22}), \qquad R_2 = R_{21} + R_{22}$$
(A~21)

In the sequel, we prove if the following constraints hold:

$$\begin{aligned} R_{e,1} &< R_1 && (a)\\ R_{e,2} &< R_{21} && (b)\\ R_1 &< I(U;Y_1) - I(U;X_2) && (c)\\ R_{21} &< I(V;Y_2|X_2) && (d)\\ R_{22} &< I(X_2;Y_2) && (e)\\ R_1 + R_{21} &< I(U;Y_1) + I(V;Y_2|X_2) - I(U;V,X_2) && (f) \end{aligned}$$
(A~22)

then $(R_1, R_2, R_{e,1}, R_{e,2})$ is achievable. Note that considering (A~20) and (A~22), and by applying a simple Fourier-Motzkin elimination to remove $R_{21}$ and $R_{22}$ from the bounds (A~22), we obtain the desired rate region. Define:

$$\begin{aligned} L_1 &\triangleq \min\{I(U;Y_1) - I(U;Y_2,V,X_2), R_1\} \\ L_1^b &\triangleq I(U;Y_2,V,X_2) - O(\epsilon) \\ L_{21} &\triangleq \min\{I(V;Y_2|X_2) - I(V;Y_1,U|X_2), R_{21}\} \\ L_{21}^b &\triangleq I(V;Y_1,U|X_2) - O(\epsilon) \end{aligned}$$
(A~23)

where $O(\epsilon) \to 0$ as $\epsilon \to 0$. Based on the definitions (A~23) and using (A~22; c) and (A~22; d), one can show:

$$\begin{aligned} R_1 &\leq L_1 + L_1^b < I(U;Y_1) \\ R_{21} &\leq L_{21} + L_{21}^b < I(V;Y_2|X_2) \end{aligned}$$
(A~24)

*Random Codebook Generation:*

1. Generate at random $2^{nR_{22}}$ independent codewords $X_2^n$ according to $P(x_2^n) = \prod_{t=1}^n P_{X_2}(x_{2,t})$. Label these codewords as $X_2^n(m_{22})$ where $m_{22} \in [1:2^{nR_{22}}]$.
2. For each $X_2^n(m_{22})$ randomly generate $2^{n(L_{21}+L_{21}^b)}$ independent codewords $V^n$ according to $P(v^n) = \prod_{t=1}^n P_{V|X_2}(v_t|x_{2,t})$. Label these codewords as $V^n(m_{22}, m_{21}, l_{21})$ where $m_{21} \in [1:2^{nR_{21}}]$ and $l_{21} \in \left[1:2^{n(L_{21}+L_{21}^b-R_{21})}\right]$.
3. Generate at random $2^{n(L_1+L_1^b)}$ independent codewords $U^n$ according to $P(u^n) = \prod_{t=1}^n P_U(u_t)$. Label these codewords as $U^n(m_1, l_1)$ where $m_1 \in [1:2^{nR_1}]$ and $l_1 \in \left[1:2^{n(L_1+L_1^b-R_1)}\right]$.
4. For each triple $\left(X_2^n(m_{22}), V^n(m_{22}, m_{21}, l_{21}), U^n(m_1, l_1)\right)$, randomly generate a codeword $X_1^n$ according to $P(x_1^n) = \prod_{t=1}^n P_{X_1|UVX_2}(x_{1,t}|u_t, v_t, x_{2,t})$. Label this codeword as $X_1^n(m_{22}, m_{21}, l_{21}, m_1, l_1)$.

*Encoding:* Given a triple $(m_1, m_{21}, m_{22}) \in [1:2^{nR_1}] \times [1:2^{nR_{21}}] \times [1:2^{nR_{22}}]$, the primary transmitter transmits $X_2^n(m_{22})$. Now consider the following set:

$$\left\{\begin{matrix} (l_{21}, l_1) \in \left[1:2^{n(L_{21}+L_{21}^b-R_{21})}\right] \times \left[1:2^{n(L_1+L_1^b-R_1)}\right]: \\ \left(X_2^n(m_{22}), V^n(m_{22}, m_{21}, l_{21}), U^n(m_1, l_1)\right) \in \mathcal{T}_\epsilon^n\left(P_{X_2VU}\right) \end{matrix}\right\}$$
(A~25)

If nonempty, the cognitive transmitter randomly uniformly chooses an element $(l_{21}^T, l_1^T)$ from this set and transmits $X_1^n(m_{22}, m_{21}, l_{21}^T, m_1, l_1^T)$. Otherwise, if the set given in (A~25) is empty, the cognitive transmitter declares encoding error and transmits an arbitrary codeword $X_1^n$.

We remark that in the coding scheme given above we have applied a novel single phase binning in which the sizes of the bins have been intelligently designed to simultaneously guarantee both jointly typical encoding (nonemptyness of the set (A~25)) and preserving of confidentiality of messages. Compared to the double binning technique given in [18] and [19] for the broadcast channel with confidential messages, ours requires a simpler scheme for codeword assigning to messages.

For bounding the probability of encoding error, we exploit the multivariate covering lemma below.

*Lemma 2)* Consider a joint PDF $P_{X_2VU}(x_2, u, v)$ and its marginal PDFs $P_{X_2}(x_2)$, $P_{V|X_2}(v|x_2)$ and $P_U(u)$. Let $0 < \epsilon_1 < \epsilon_2 < p_{min}(P_{X_2VU})$ where $p_{min}(P_{X_2VU})$ denotes the minimum positive value of $P_{X_2VU}$. Also, let $(B_1,B_2)\in\mathbb{R}_+^2$ be a pair of non-negative real numbers. Given the sequence $x_2^n\in\mathcal{T}_{\epsilon_1}^n(P_{X_2})$, a random codebook is generated as follows:

1. Randomly generate $2^{nB_1}$ independent codewords $U^n$ according to $P(u^n)=\prod_{t=1}^n P_U(u_t)$ . Label these codewords as $U^n(b_1)$ where $b_1\in[1:2^{nB_1}]$.
2. For the given $n$-sequence $x_2^n$, randomly generate $2^{nB_2}$ independent codewords $V^n$ according to $P(v^n)=\prod_{t=1}^n P_{V|X_2}(v_t|x_{2,t})$. Label these codewords as $V^n(x_2^n,b_2)$ where $b_2\in[1:2^{nB_2}]$.

Then, there exists $O(\epsilon)\to 0$ as $\epsilon\to 0$, such that if:

$$\begin{cases} B_1 > I(U;X_2)+O(\epsilon) \\ B_1+B_2 > I(U;V,X_2)+O(\epsilon)\end{cases}$$

we have:

$$P\left(\bigcap_{\substack{b_i,i=1,2\\ b_i\in[1:2^{nB_i}]}} \left(V^n(x_2^n,b_2),U^n(b_1)\right)\notin\mathcal{T}_{\epsilon_2}^n\left(P_{X_2VU}|x_2^n\right)\middle| x_2^n\right)\xrightarrow{n\to\infty}0$$

*Proof of Lemma 2)* This result is derived by setting $U_0\equiv W_1\equiv\emptyset$ and $B_0=0$ and also by redefining $U_1\cong U$, $V_1\cong V$ and $V_0\cong X_2$ in [28, pp. 10-11, Lemma 1]. ■

Now based on Lemma 2, we can deduce that the error probability of encoding vanishes provided that:

$$\begin{gathered} L_1+L_1^b-R_1>I(U;X_2)\\ L_1+L_1^b-R_1+L_{21}+L_{21}^b-R_{21}>I(U;V,X_2)\end{gathered}\tag{A~26}$$

Considering the definitions (A~23), one can show that (A~22; c) and (A~22; f) guarantee the constraints (A~26).

*Decoding:*

1. At the cognitive receiver, the decoder tries to find a unique pair $(\hat{m}_1,\hat{l}_1)$ with:

   $$\left(U^n(\hat{m}_1,\hat{l}_1),Y_1^n\right)\in\mathcal{T}_\epsilon^n(P_{UY_1})$$

   If there exists such a pair and it is unique, the decoder estimates its respective message by the corresponding $\hat{m}_1$. Otherwise, a decoding error is declared.
2. At the primary receiver, the decoder tries to find a unique triple $(\hat{m}_{22},\hat{m}_{21},\hat{l}_{21})$ with:

   $$\left(X_2^n(\hat{m}_{22}),V^n(\hat{m}_{22},\hat{m}_{21},\hat{l}_{21}),Y_2^n\right)\in\mathcal{T}_\epsilon^n(P_{X_2VY_2})$$

   If there exists such a triple and it is unique, the decoder estimates its respective message by the corresponding $(\hat{m}_{22},\hat{m}_{21})$. Otherwise, a decoding error is declared.

One can readily show that the constraints (A~24) and also (A~22; e) guarantee the decoding error probability at both receivers vanishes.

*Equivocation Calculation:* Now we derive the bounds on the equivocation rates (A~20). To evaluate the equivocation rate of $M_2$ at the receiver $Y_1$, we can write:

$$\begin{aligned} &H(M_{21},M_{22}|Y_1^n)\\ &\geq H(M_{21}|Y_1^n,U^n,X_2^n,M_{22})\\ &= H(M_{21},Y_1^n|U^n,X_2^n,M_{22})-H(Y_1^n|U^n,X_2^n,M_{22})\\ &= H(M_{21},Y_1^n,V^n|U^n,X_2^n,M_{22})\\ &\quad -H(V^n|Y_1^n,U^n,X_2^n,M_{21},M_{22})-H(Y_1^n|U^n,X_2^n,M_{22})\\ &= H(M_{21},V^n|U^n,X_2^n,M_{22})+H(Y_1^n|U^n,V^n,X_2^n,M_{21},M_{22})\\ &\quad -H(V^n|Y_1^n,U^n,X_2^n,M_{21},M_{22})-H(Y_1^n|U^n,X_2^n,M_{22})\\ &\overset{(a)}{\geq} H(V^n|U^n,X_2^n)+H(Y_1^n|U^n,V^n,X_2^n)\\ &\quad -H(V^n|Y_1^n,U^n,X_2^n,M_{21},M_{22})-H(Y_1^n|U^n,X_2^n)\\ &= H(V^n|X_2^n)-I(V^n;U^n|X_2^n)\\ &\quad -I(V^n;Y_1^n|U^n,X_2^n)-H(V^n|Y_1^n,U^n,X_2^n,M_{21},M_{22})\end{aligned}\tag{A~27}$$

where inequality (a) holds because given $X_2^n$ the message $M_{22}$ is uniquely determined, conditioning does not increase the entropy, and also $M_{21},M_{22}\to(U^n,V^n,X_2^n)\to Y_1^n$ forms a Markov chain. Now consider the first term in the right hand side of (A~27). According to the codeword generation, conditioned on $X_2^n$, $V^n$ has $2^{n(L_{21}+L_{21}^b)}$ possible values with equal probability. Thus,

$$H(V^n|X_2^n)=\log 2^{n(L_{21}+L_{21}^b)}=n(L_{21}+L_{21}^b)\tag{A~28}$$

Also, for the second and the third terms in (A~27) using the same approach as [18, Lemma 3], one can show:

$$\begin{gathered} I(V^n;U^n|X_2^n)\leq nI(V;U|X_2)+nO(\epsilon)\\ I(V^n;Y_1^n|U^n,X_2^n)\leq nI(V;Y_1|U,X_2)+nO(\epsilon)\end{gathered}\tag{A~29}$$

Then, consider the last term in (A~27). Given $U^n,X_2^n$, and $(M_{21},M_{22})=(m_{21},m_{22})$, let $P_e^{Y_1^n}(m_{21},m_{22})$ denotes the error probability for receiver $Y_1$ to decode $V^n$. We show that $P_e^{Y_1^n}(m_{21},m_{22})\to 0$ as $n\to 0$. A decoding strategy is adapted as follows. The receiver $Y_1$ tries to find a unique $\hat{l}_{21}$ with:

$$\left(V^n(m_{22},m_{21},\hat{l}_{21}),U^n,X_2^n,Y_1^n\right)\in\mathcal{T}_\epsilon^n(P_{VUX_2Y_1})$$

If there is no such $\hat{l}_{21}$ or there are more than one, a decoding error is declared. By a standard analysis of error probability, one can show $P_e^{Y_1^n}(m_{21},m_{22})\leq O(\epsilon)$ provided that:

$$L_{21}+L_{21}^b-R_{21}\leq I(V;Y_1,U|X_2)-O(\epsilon)\tag{A~30}$$

Note that the constraint (A~30) is guaranteed by the definitions (A~23). Thereby, we can write:

$$\begin{aligned} &H(V^n|Y_1^n,U^n,X_2^n,M_{21},M_{22})\\ &= \sum\nolimits_{m_{21},m_{22}}P(m_{21},m_{22})H(V^n|Y_1^n,U^n,X_2^n,m_{21},m_{22})\\ &\overset{(a)}{\leq} \sum\nolimits_{m_{21},m_{22}}P(m_{21},m_{22})\begin{pmatrix}1+P_e^{Y_1^n}(m_{21},m_{22})\\ \times\log 2^{n(L_{21}+L_{21}^b-R_{21})}\end{pmatrix}\\ &\leq \left(1+O(\epsilon)\times n(L_{21}+L_{21}^b-R_{21})\right)\end{aligned}\tag{A~31}$$

where (a) is due to the Fano's inequality. Now by substituting (A~28), (A~29), and (A~31) in (A~27) we derive:

$$\frac{1}{n}H(M_2|Y_1^n) \geq L_{21} + L_{21}^b - I(V;U|X_2) - I(V;Y_1|U,X_2)$$
$$-O(\epsilon) - \left(\frac{1}{n} + O(\epsilon) \times (L_{21} + L_{21}^b - R_{21})\right)$$
$$= L_{21} - O(\epsilon)$$

Therefore, $R_{e,2}$ is achievable if:

$$R_{e,2} \leq L_{21} \quad \text{(A~32)}$$

The constraints (A~20) and (A~22; b) assure us that (A~32) holds. By following rather similar steps, one can derive that the equivocation $R_{e,1}$ is achievable provided that:

$$R_{e,1} \leq L_1 \quad \text{(A~33)}$$

which is guaranteed by (A~20) and (A~22; a). The proof is thus complete. ■

***Proof of Theorem 7)*** To derive the direct part, consider the achievable rate region (10) for the case where no confidentiality is required for the primary message $M_2$. Let $W$ and $\tilde{X}_1$ be independent Gaussian RVs with zero means and unit variances. Let also $\alpha$ be an arbitrary real number belonging to the interval [0,1]. Define:

$$\begin{cases} V \equiv Q \equiv \emptyset \\ X_2 \equiv \sqrt{P_2}W \\ X_1 \equiv \sqrt{\alpha P_1}\tilde{X}_1 + \frac{|b|}{b}\sqrt{(1-\alpha)P_1}W \\ U \equiv \tilde{X}_1 + \frac{\sqrt{\alpha P_1}\left(\frac{|b|}{b}\sqrt{(1-\alpha)P_1} + a\sqrt{P_2}\right)}{\alpha P_1 + 1}W \end{cases}$$

Now, by substituting $W, X_1, X_2$, and $U$ in (10), we have:

$$\begin{cases} I(U;Y_1) - I(U;W,X_2) = I(U;Y_1|W,X_2) = \psi(\alpha P_1) \\ , \\ I(W,X_2;Y_2) = \psi\left(\frac{(1-\alpha)b^2P_1 + P_2 + 2|b|\sqrt{(1-\alpha)P_1P_2}}{\alpha b^2 P_1 + 1}\right) \\ , \\ I(U;Y_1) - I(U;Y_2,W,X_2) = I(U;Y_1|W,X_2) - I(U;Y_2|W,X_2) \\ \qquad = \psi(\alpha P_1) - \psi(\alpha b^2 P_1) \end{cases}$$

Therefore, the rate region (20) is achievable. To prove the converse part, using Theorem 1 we can deduce that the following rate region constitutes a valid outer bound for the case where only the message $M_1$ (see Fig. 1) is confidential:

$$\bigcup_{P_{UVX_1X_2}} \left\{\begin{matrix} (R_1, R_2, R_{e,1}) \in \mathbb{R}_+^3 : \\ R_{e,1} \leq R_1 \\ R_1 \leq I(U,V;Y_1|X_2) \\ R_2 \leq I(V,X_2;Y_2) \\ R_1 + R_2 \leq I(U;Y_1|V,X_2) + I(V,X_2;Y_2) \\ R_{e,1} \leq [I(U;Y_1|V,X_2) - I(U;Y_2|V,X_2)]_+ \end{matrix}\right\} \quad \text{(A~34)}$$

Now, by following the same arguments as the proof of Theorem 3, one can derive that for the Gaussian channel (1) with $|b| \leq 1$, the rate region (A~34) is equivalent to (20). Note that, in Theorem 3, the condition $ab = 1$ is only required to prove $R_{e,2} = 0$; however, in Theorem 7, we are not interested in measuring the confidentiality of the primary message $M_2$. The proof is thus complete. ■